# THE CONCEPT, TYPES AND STRUCTURE OF CORRUPTION

**Oleg Antonov,**
*PhD in Law, Associate Professor*
*Law Institute of the Russian University of Transport, Russia, Moscow*
**Ekaterina Lineva,**
*PhD in Law, Associate Professor*
*Law Institute of the Russian University of Transport, Russia, Moscow*

**Abstract**

The article analyzes the essence of the phenomenon of corruption, highlights its main varieties and characteristics. The authors of the study apply historical analysis, emphasizing the long-term nature of corruption and its historical roots. The paper uses legal analysis to characterize the legal interpretation of corruption as an economic crime.

## 1. Introduction

The existence of corruption and its negative consequences have attracted the attention of scientists for many decades, but in recent years, since the 1990s, the so-called 'eruption of corruption' has occurred in the world (Jiang, 2017), which has forced researchers to further address this issue. The exact time of the appearance of corruption is quite difficult to determine, but we can assume with a high degree of confidence that this phenomenon had been formed in connection with the division of society into more and less prosperous classes. References to corruption can be found in the texts of the Ancient Sumer, in the treatises of the Ancient Egypt and the Ancient India, as well as in the texts of the Ancient Greek and Roman philosophers (Kuzovkov, 2010).

It should be noted that more ancient references characterize corruption rather as an offering of gifts to 'powerful' persons, whereas in the Ancient Rome the word 'corrumpere' itself is already used, meaning 'to corrupt', 'to violate', and 'to break' (Macievsky, 2014). Simultaneously with the appearance of the phenomenon of corruption, the authorities began to fight it: in the Ancient Rome, the punishment for corruption was the death penalty. Thus, corruption had developed and changed along with the history of a mankind: in the Middle Ages, the phenomenon had been already well known in the public environment, and by the end of the XIX-XX centuries it has already spread widely in the economic and political spheres.

It is noteworthy that in Russia the phenomenon of corruption had been also widely known, starting from the Middle Ages, when Metropolitan Kirill had called for the death of corrupt officials (Fedorina, 2014) according to the ancient Roman tradition. The provisions of the Dvina Charter of 1397-1398 indicate such a phenomenon as a promise-illegal remuneration for the exercise of official powers in Russian legislation (Avanesyan, 2013), which suggests that bribery had been already widespread at that time. It can also be noted that nowadays the concept of corruption includes what in Russia previously had been called as 'bribery' and 'embezzlement' (Permyakov, 2010), which further confirms the broad meaning and application of the term under study.

However, identifying a common, universally accepted understanding of corruption leads to difficulties in the academic environment. Some researchers are trying to identify certain behaviors that may be the subject of attention, others are looking for the cultural aspect of corruption, and still others represent the scenario of behavior within the framework of corruption. In addition, existing definitions of corruption are often problematic, as they are either too extensive, making them relatively useless, or too narrowly focused, which automatically makes them rarely applicable and only in certain cases.

## 2. Main part

The problem of defining corruption is related to the fact that, in general, corruption is a process that accompanies any exercise of state power and does not belong to any particular type of political regime. It is also not related to the development of any specific country, but can be applied to a very wide range of human behavior (Afanasyeva et al., 2019), as well as to state institutions at different levels. Based on this, it can be assumed that it is impossible to develop only one generalized and indisputable definition of political corruption, which is reflected in the numerous forms of bribery, the relativity of norms and law, as well as in the ambiguous boundaries between public and private law.

In accordance with the Russian legislature (Federal Law FZ-273, 2008), corruption means abuse of official position, giving and receiving bribes, abuse of authority, commercial bribery or other illegal use by an individual of his/her official position contrary to the legitimate interests of society and the state in order to obtain benefits in the form of money, valuables, other property or services of a property nature, other property rights for himself/herself or for third parties, or illegal provision of such benefits to the specified person by other individuals; as well as the commission of these acts on behalf of or in the interests of a legal entity.

The Norwegian researcher Amundsen offers a whole set of definitions of corruption (Amundsen, 2000):

- this is a cash payment in a transitive relationship;
- a misappropriation of public funds;
- an economic fraud or deception;
- extracting money or other resources from disadvantaged individuals through coercion, threats, or even violence;
- natural human inclinations in favor of friends, relatives, or someone close and trusted in the process of resource allocation or in the political process.

At the same time, in the Explanatory Dictionary of the Russian Language (Ozhegov's Explanatory Dictionary), the Big Encyclopedic dictionary, the Russian Legal Encyclopedia, the UN Reference Document on the International Fight against Corruption (UN Declaration, 1996), the concept of corruption is considered as a set of acts (crimes, offenses) of public servants (officials) that consist in the selfish use of their power (official rights and powers, position, opportunities, status) for personal gain, for the benefit of third parties or groups, and leading to moral corruption in the power structures.

As a result, corruption is most generally understood as abuse of power for personal gain, as well as the bribery of individuals and their venality (Fedorina, 2014). Thus, corruption usually refers to the use by an official of his authority and the rights entrusted to him for personal gain, which is contrary to the rules established by law. In this sense, corruption will be considered in the course of this study.

Like identifying the definition of corruption, its typology is also a challenge: the research literature is full of a variety of corruption classifications. For instance, according to the scale of the crime, corruption is divided into large and small types. Thus, corruption on a particularly large scale occurs at the highest levels of the political system (Kuzovkov, 2010), such as the participation of the president, ministers and other senior officials, and therefore involves significant payments. Small-scale corruption occurs at low levels, such as administrative divisions where citizens apply for services (Kuzovkov, 2010). In turn, minor corruption can be divided into cases of separate corruption, official and institutional, which is carried out directly in favor of the interests of officials.

There is also a distinction between 'systemic' and 'systematic' corruption. Systemic corruption is a product of weak administrative systems and a lack of control institutions, thus affecting the system (Sheverdyaev, 2016). On the contrary, systematic corruption, considered as organized crime, occurs when activities that pursue private interests undermine the very structure of the political system or can be seen as the result of political manipulation of economic interests (Rimsky, 2004). In any case, the problem of institutional corruption seems to be one of the main threats to state security.

In this regard, it can be concluded that the versatility of the application of the term 'corruption' has led to the existence of a number of classifications. In addition, the diversity of approaches to the study of corruption can be explained by the spread of this phenomenon in various spheres of human activity, which has revealed the need for its study by representatives of various specialties (economists, political scientists, sociologists, psychologists, lawyers). In turn, this can somehow manifest itself in the one-dimensionality of the provided classifications and the presentation of only a few signs of the phenomenon under study (Akhmetova, 2008), which does not claim to comprehensively reflect the types of acts of corruption.

The researchers classify corruption according to the degree of participation of authorized persons in the distribution of resources into civilized and political. Progressive or civilized corruption assumes that officials indirectly participate in the distribution of profits by entrepreneurs (Akhmedov, 2003), and the state recognizes the phenomenon of corruption through its resolution (Akhmetova, 2008). In this case, political corruption is understood as an act in which state and public figures and politicians are the participants (Verbin, 2003). Within the framework of political corruption, attention is drawn to the allocation of corruption in the field of public administration, parliamentary corruption, acts of corruption in enterprises and during elections (Proyava, 2012). In these cases, the actions are aimed at the possible use of the resources of civil servants and influence on their decision-making, as well as promoting the interests of selected groups in the adoption of legislative norms or vote buying during the elections (Zyryanova, 2014).

Corruption activities can also be distinguished by the types of socio-economic relations, which allows researchers to divide the nature of corruption into 'western' and 'eastern' forms (Dobrenkov, 2009). Thus, the 'Western' corruption should be understood as a kind of market for corrupt services, in which the parties enter into a one-time purchase and sale relationship. In other words, the 'Eastern' corruption denotes a certain system of regular relations, closely connected with such social ties as family, professional, and corporate relations.

In any case, the most attractive objects of corruption acts, as a rule, are state projects and tenders, business activities of relatives of officials, state mechanisms of institutional influence (Kuzminov, 2002). It seems logic that the socio-economic sphere faces the most frequent manifestations of corruption. The corruption schemes also affect shadow areas of the economy.

In order to generalize the existing types of typologies, the phenomenon of corruption can be divided into four types, based on the field of activity: by subject composition, by the object of influence, by the nature of the impact on the regulated relations, and by its scope or size (Golubovsky, 2015).

According to the object of influence, corruption can spread:
- in the executive authorities;
- in the legislative branch;
- in the judiciary;
- in local government bodies;
- in commercial and non-profit organizations.

According to the nature of the impact on the regulated relations, an act of corruption may be followed by:
- civil law violations;
- criminal offences;
- administrative and disciplinary offenses;
- abuse of public status (nepotism, lobbying, protectionism, providing official information to third parties).

By its scope or size, corruption can manifest itself (Laptev et al., 2014):
- at the national level, operating within the same country and the same legal system;
- at the regional level;
- at the international (inter-national) level, covering several states at once.

Accordingly, the structure of corruption can be determined by the subject composition (Satarov, 2013):

- high-level corruption involving politicians and government officials at the federal level, involving the adoption of laws and decisions at the highest level;
- grass-roots corruption, which manifests itself in the direct interaction of a person with representatives of local self-government bodies;
- special corruption that occurs in non-governmental organizations

In other words, corruption can be structured in hierarchical way, when lower-ranking officials corrupt higher-ranking officials (Zenyuk, 2016). Such a model of 'bottom-up' corruption behavior is more common than the reverse, due to the fact that the upper levels of power institutions and the resources in their possession are attractive to the initiators of the act of corruption.

## 3. Conclusion

Thus, corruption can be defined as acts of bribery that are provided for the purpose of influencing another person in the interests of the corrupt official. Corruption allows you to benefit through illegal actions and at the same time destroy the system of which corrupt officials are a part. Corruption can occur on different scales: it ranges from small indulgences between a small number of people to large-scale acts of corruption affecting the Government.

As a result, corruption has become so widely spread that today it could be considered as a part of the everyday structure of society, including corruption as one of the symptoms of organized crime. In this regard, corruption can occur in any sector, whether private or public, but the public interest in developing internal mechanisms to combat active or passive corruption arises mainly only in institutions under the control of democratic structures.

10. Corruption // Explanatory dictionary of the Russian language by S. Ozhegov, N. Shvedova. [Electronic data]. Available from: http://ozhegov.info/slovar/?ex=Y&q=CORRUPTION

11. Corruption // Big Encyclopedic dictionary [Electronic data]. Available from: https://dic.academic.ru/dic.nsf/enc3p/164642

12. Corruption // Russian Legal Encyclopedia [Electronic data]. Available from: https://yuridicheskaya_encyclopediya.academic.ru/4873

13. United Nations Declaration against Corruption and Bribery in International Commercial Transactions: adopted by General Assembly resolution 51/191 on December 16, 1996. [Electronic data]. Available from: http://www.un.org/ru/documents/decl_conv/declarations/bribery.shtml

14. Sheverdyaev, S. N. (2016) System corruption as a problem of the science of constitutional law: discussion of the issue at the Faculty of Law of Lomonosov Moscow State University // Constitutional and Municipal Law. No. 9. pp. 10-16.

15. Rimsky, V. L. (2004) Bureaucracy, clientilism and corruption in Russia // Social Sciences and Modernity. No. 6. pp. 68-79.

16. Akhmetova, N. A. (2008) The social mechanism of corruption reproduction in the conditions of modern Russian society. Volgograd: Volga. - p. 45.

17. Akhmedov, I. (2003) Corruption lawlessness // Monitor. No. 14 (41) [Electronic data]. Available from: http://transparency.az/transpfiles/zig28.pdf

18. Verbin, A. (2003) Ne podmazhesh, ne podedesh [If you don't give me a gift, you won't be able to approach me] // Soviet Russia (in Russian).

19. Proyava, S. M. (2012) Economization of corruption. The mechanism of counteraction. M.: UNITY. - Pp. 32-33.

20. Zyryanova I. A. (2014) Corruption in the electoral process: concept and signs // Criminal Justice. No. 1 (3). pp. 97-100.